\documentclass[12pt]{article}

\usepackage[T1]{fontenc}
\usepackage[utf8]{inputenc}
\usepackage{amsfonts,amsmath,amssymb,accents,mathrsfs}
\usepackage[retainorgcmds]{IEEEtrantools}
\usepackage{multirow}
\usepackage{multicol}
\usepackage{tikz}
\usepackage{graphicx,color}
\usepackage{booktabs}
\usepackage{placeins}
\usepackage{mathtools}
\usepackage{calc}
\usepackage{youngtab}
\usepackage[nosort,nomove,compress]{cite}
\usepackage[scaled=0.9]{helvet}
\usepackage[letterpaper,%showframe,
            margin=2.282cm]{geometry}
\usepackage{setspace}
\usepackage[colorlinks,
			breaklinks,
			hyperfootnotes,
			linktoc=page,
            linkcolor=blue,
            citecolor=blue,
            urlcolor=blue,
            unicode,
            pdfstartview=FitH,
            bookmarks,
            bookmarksnumbered]{hyperref}

\renewcommand\a {{\alpha}}
\renewcommand\b {{\beta}}
\newcommand\g {{\gamma}}
\renewcommand\d {{\delta}}
\renewcommand\r {{\rho}}

\newcommand\e {{\epsilon}}

\renewcommand\L {{\Lambda}}
\renewcommand\l {{\lambda}}

\newcommand\ad {{\dot{\alpha}}}
\newcommand\bd {{\dot{\beta}}}
\newcommand\gd {{\dot{\gamma}}}

\newcommand\rd {{\dot{\rho}}}
\newcommand\sd {{\dot{\sigma}}}

\newcommand\Ysf{{\textsf{Y}}}
\newcommand\N{{\mathcal{N}}}
\newcommand\E{{\mathcal{E}}}
\newcommand\I{{\mathcal{I}}}

\newcommand\J{{\mathcal{J}}}
\renewcommand\H{{\mathcal{H}}}
\newcommand\V{{\mathcal{V}}}
\newcommand\W{{\mathcal{W}}}
\newcommand\X{{\mathcal{X}}}
\newcommand\vO{{\varOmega}}

\newcommand\D {{\rm D}}
\newcommand\Dd {{\bar{\rm D}}}
\newcommand\pa {{\partial}}

\def\bea{\begin{IEEEeqnarray*}}
\def\eea{\end{IEEEeqnarray*}}
\def\be{\begin{eqnarray}}
\def\ee{\end{eqnarray}}
\def\n{\IEEEyesnumber}
\def\sn{\IEEEyessubnumber}

\makeatletter
\renewcommand\section{\@startsection{section}{1}{\z@}
              {3ex plus-1ex minus-.2ex}{1pt plus1pt}
              {\large\sf\bfseries\boldmath}}
\renewcommand{\subsection}{\@startsection{subsection}{2}{\z@}
              {1.5ex plus-1ex minus-.2ex}{0.01pt plus1pt}{\sf\slshape}}
\renewcommand{\subsubsection}{\@startsection{subsubsection}{3}{\z@}
              {1.5ex plus-1ex minus-.2ex}{0.01pt plus0.2pt}{\sf\boldmath}}
\renewcommand{\paragraph}{\@startsection{paragraph}{4}{\z@}
              {.75ex \@plus.5ex \@minus.2ex}{-2mm}{\sf\bfseries\boldmath}}
\makeatother

\newcommand{\Title}[1]{ {\large \bf #1 \vspace{3ex}} }
\newcommand{\Author}[3]{ {\large #1\footnote{\href{mailto:#2}{#2}}$^{#3}$} }
\newcommand{\Inst}[2]{ \emph{\centering $^{#2}$#1} }
\newcommand{\Abstract}[1]{ { ABSTRACT}\\ [4mm]
  \parbox{142mm}{\parindent=2pc\indent\baselineskip=14pt plus1pt #1}}

\begin{document}
\thispagestyle{empty}
\vspace*{6mm}
\begin{center}
    \Title{Superspace First Order Formalism, Trivial Symmetries and\\[5pt]
Electromagnetic Interactions of Linearized Supergravity} \\   [9mm]
% \Author{author1}{author1 email address}{affiliation index}~,~
\Author{I.~L.~Buchbinder}{joseph@tspu.edu.ru}{a,b},~%\\ [8mm]
\Author{S.~James Gates Jr.}{sylvester\_gates@brown.edu}{c,d},~%\\ [8mm]
\Author{K.~Koutrolikos}{konstantinos\_koutrolikos@brown.edu}{c,d}\\ [8mm]
\Inst{Center of Theoretical Physics,Tomsk State Pedagogical University,\\
      5 Tomsk 634041, Russia}{a}\\[8pt]
\Inst{National Research Tomsk State University,\\
      8 Tomsk 634050, Russia}{b}\\[8pt]
\Inst{Brown Theoretical Physics Center,\\[1pt]
      Box S, 340 Brook Street, Barus Hall,
      Providence, RI 02912, USA}{c}\\[8pt]
\Inst{Department of Physics, Brown University,\\[1pt]
      Box 1843, 182 Hope Street, Barus \& Holley 545,
      Providence, RI 02912, USA}{d}\\ [10mm]
\Abstract{
We introduce a first order description of linearized non-minimal ($n=-1$) supergravity in superspace, using
the unconstrained prepotential superfield instead of the conventionally constrained super one forms.  In this
description, after integrating out the connection-like auxiliary superfield of first-order formalism, the
superspace action is expressed in terms of a single superfield which combines the prepotential and compensator
superfields. We use this description to construct the supersymmetric cubic interaction vertex $3/2-3/2-1/2$
which describes the electromagnetic interaction between two non-minimal supergravity multiplets (superspin
$\Ysf=3/2$ which contains a spin 2 and a spin 3/2 particles) and a vector multiplet (superspin $\Ysf=1/2$
contains a spin 1 and a spin 1/2 particles). Exploring the trivial symmetries emerging between the two
$\Ysf=3/2$ supermultiplets, we show that this cubic vertex must depend on the vector multiplet superfield
strength. This result generalize previous results for non-supersymmetric electromagnetic interactions of spin
2 particles. The constructed cubic interaction generates non-trivial deformations of the gauge
transformations.
}
\end{center}
\vfill
%
% \Keywords{--keywords go here--}
% \vfill
\clearpage
%
% \tableofcontents
%%%%%%%%%%%%%%%%%%%%%%%%%%%%
%%%%%%%%%%%%%%%%%%%%%%%%%%%%
\section{Introduction}
\label{sec:intro}
%%%%%%%%%%%%%%%%%%%%%%%%%%%%
%%%%%%%%%%%%%%%%%%%%%%%%%%%%
It has been understood that in the high energy regime ($\a'\to\infty$) of string theory, the role of higher
spin modes is crucial. The unifying framework of the theory, all string excitations must be treated equally,
strongly suggests that we should seriously consider higher spin fields and their interactions (preferably with
supersymmetry)  and overcome the discouragement of early no-go theorems.
The investigation of consistent cubic interactions among higher spin fields is a very important first step in
this direction. The classification of all non-supersymmetric cubic vertices of massive and massless integer
and half-integer higher spin fields  was obtained by Metsaev in
\cite{Metsaev:2005ar,Metsaev:2007rn,Metsaev:2018xip} for
$d\geq4$ using light-cone formalism and in \cite{Mkrtchyan:2017ixk,Kessel:2018ugi} for $d=3$. More recently a
general classification of supersymmetric higher spin multiplets was also found in
\cite{Metsaev:2019dqt,Metsaev:2019aig}.

The covariant construction of such cubic interactions is a highly
non-trivial task and most of the successful constructions involve massless fields. In these cases, the gauge
redundancy of the higher spin fields is used as a guiding principle which severely constraints the
structure of the interaction vertex. A variety of methods have been used to construct such vertices
(Noether's procedure~\cite{Berends:1985xx,Barnich:1993vg,Bekaert:2009ud,Bekaert:2010hk,Buchbinder:2012iz,
Joung:2012fv,Joung:2013nma},
BRST~\cite{Buchbinder:2006eq,Bekaert:2006us,Fotopoulos:2007yq,Fotopoulos:2008ka,Polyakov:2009pk,
Henneaux:2012wg,Metsaev:2012uy,Henneaux:2013gba,Buchbinder:2021qrg,Buchbinder:2021xbk},
frame-like formulation~\cite{Vasiliev:1980as,Fradkin:1987ks,Lopatin:1987hz,Vasiliev:1987tk,
    Vasiliev:2001wa,Alkalaev:2002rq,Zinoviev:2008ze,Ponomarev:2010st,Zinoviev:2010cr,
    Boulanger:2011qt,Zinoviev:2011fv,Buchbinder:2019dof,Buchbinder:2019olk,Buchbinder:2019kuh,
Buchbinder:2020rex,Khabarov:2020bgr,Khabarov:2021xts})
but all of them are organized in two categories.
There is the metric-like description, which uses symmetric, higher rank tensors
resembling the metric formulation of gravity and there is the frame-like description which is a
generalization of the vielbein description of gravity, where the higher spin fields are described by higher
spin algebra valued soldering one forms.

The metric-like formulation, originates from early works
\cite{Dirac:1936tg,Fierz:1939ix,Wigner:1939cj,Bargmann:1948ck,Singh:1974qz,Singh:1974rc,Fronsdal:1978rb,
Fang:1978wz},
offers a more geometric viewpoint which together with the notion of higher spin
connections\cite{deWit:1979sib} tries to extend our spin 2 intuition to higher spins. This  description is
very economic in the number of fields it requires.
On the other hand the
frame formulation, developed by Vasiliev\cite{Vasiliev:1980as,Vasiliev:1986td}, offers a more algebraic
approach by extending the notion of Cartan connection in gravity to higher spins and thus resembling the
structure of non-abelian Yang-Mills theory for an appropriate underlying higher spin symmetry group. This
approach provides an economy of ideas which dictate the dynamics of higher spins and so far
it has been more successful in constructing consistent interactions.

For manifestly $N=1$ supersymmetric theories in 4D,
% the superspace description of higher spin multiplets uses higher
% rank superfields independently symmetrized in both types of spinorial indices. The superspace lagrangian
the superspace description of free massless higher spins was discovered in
\cite{Kuzenko:1993jq,Kuzenko:1993jp,Kuzenko:1994dm,Gates:1996my} (see also later discussions in
\cite{Gates:2013rka,Gates:2013ska,Koutrolikos:2015lqa}) whereas the description of massive higher
spin multiplets was only recently discovered in \cite{Koutrolikos:2020tel} for arbitrary half-integer
superspins (\Ysf=s+1/2). These formulations use higher rank superfields with both types of spinorial indices
symmetrized. They can be understood as the
superspace analogs to the geometrical, metric-like approach (see the discussion in
\cite{Buchbinder:2020yip}) and have been used successfully to construct interactions of higher
spin supermultiplets. In \cite{Buchbinder:2017nuc,Hutomo:2017phh,Hutomo:2017nce,Koutrolikos:2017qkx,
Buchbinder:2018wwg,Buchbinder:2018nkp,Buchbinder:2018gle,Hutomo:2018tjh} various non-trivial cubic
interactions between higher spin multiplets and matter multiplets have been constructed. Cubic interactions
$\Ysf_1-\Ysf_2-\Ysf_3$ among higher spin multiplets with superspins $\Ysf_1, \Ysf_2, \Ysf_3$ were first
constructed in \cite{Buchbinder:2018wzq,Gates:2019cnl} for the two cases of $\Ysf_1=s_1+1/2$, $\Ysf_2=\Ysf_3$
(integer or half-integer) and $\Ysf_1=s_1$, $\Ysf_2=\Ysf_3$ (integer or half-integer). These interactions are
of the
abelian type because the superspace cubic interaction Lagrangian is of the form
$\mathcal{L}_1\sim\Phi_1~W_2~W_3$, where $\Phi_1$ is the set of superfields that describe
the higher spin supermultiplet with superspin $\Ysf_1$ and $W_2, W_3$ are the gauge
invariant superfield strengths for higher spin multiplets with superspins $\Ysf_2,\Ysf_3$ respectively.
Because of their structure these interactions do not generate corrections to the free
gauge transformations.

To make progress in the program of constructing manifestly supersymmetric cubic interactions of higher spins,
we would like to investigate interactions of the type $\mathcal{L}_1\sim\Phi_1~\Phi_2~W_3$. Such
cubic interactions have been recently been constructed for higher spin theories with on-shell supersymmetry
\cite{Khabarov:2020deh}. In general, this type of interactions generate non-trivial deformations of the gauge
transformations that may also deform the gauge algebra. An important subclass of such interactions are the
electromagnetic interactions of higher spin multiplets, \Ysf-\Ysf-1/2.  By turning on the electric charge of a
superspin $\Ysf$ we can couple it with the vector multiplet ($\Ysf=1/2$). For electromagnetic
interactions one must consider a doublet of (super)fields in order for the U(1) group
(isomorphic to SO(2)) which controls the interaction to be able to be realized\footnote{One can also
    understand the emergence of the doublet as a consequence of charge conservation on the
cubic vertex, meaning that if we turn on the electric charge for one higher spin multiplet we must also have
another one with opposite charge.}. Such interactions have been explicitly constructed for non-supersymmetric
theories in \cite{Boulanger:2008tg,Zinoviev:2008jz,Zinoviev:2009hu,Zinoviev:2010av} and motivate us to find
the corresponding manifestly supersymmetric structures.

In this paper we focus on the first member of this class of interactions, the electromagnetic interactions of
linearized supergravity ($\Ysf=3/2$): $3/2-3/2-1/2$\footnote{Not to be confused with the (super)gravitational
interaction of the vector multiplet: $3/2-1/2-1/2$.}. Specifically, we consider a doublet of linearized
non-minimal ($n=-1$) supergravity supermultiplets described by superfields $H^{i}_{\a\ad},
\chi^{i}_{\a},~i=1,2$ and a vector multiplet described by superfield $V$. By exploiting the set of trivial
symmetries of free theory, we show that the cubic interaction must depend purely on the superfield strength of
the vector multiplet $W_{\a}$ and does not include bare $V$ terms, thus making it trivially invariant under
the vector gauge transformation. On the other hand, the linearized supergravity superfields will participate
bare and via Noether's procedure a deformation of the free gauge transformations will be required.  We are
using non-minimal supergravity because, unlike the minimal formulations, it is a geometrical
theory\footnote{In the sense that its equations of motion can be written purely in terms of de
Wit-Freedman-like superconnections as found in \cite{Buchbinder:2020yip}.} and it is the lowest member of a
tower of non-minimal higher spin  supermultiplets, as observed in \cite{Kuzenko:1993jp}. Therefore any
intuition we gain by studying non-minimal supergravity will benefit us towards constructing cubic interactions
of supersymmetric higher spin multiplets.

In order to construct this interaction (and being motivated by the success of the frame-like formulation) we
develop a superspace, first order formulation of linearized supergravity by introducing a not fully symmetric
superfield $\I_{\b\a\ad}$ which is invariant under a relaxed set of gauge transformations
and combines superfields $H_{\a\ad}$ and $\chi_{\a}$. We show that the linearized non-minimal supergravity
action can be written only in terms of $\I_{\b\a\ad}$ and is the appropriate superfield
variable to construct the cubic interaction with the vector multiplet.

The paper is organized as follows. In section 2, we introduce the first order formulation of linearized
non-minimal supergravity in terms of the superfield $\I_{\b\a\ad}$ and a pair of auxiliary connection-like
superfields.  In section 3, we review the notion of trivial symmetries and give examples of their application
in supersymmetric theories. In this case the doublet of superfields gives rise to such trivial symmetries
which in turn constraint the cubic interaction to depend on the superfield strength $W_{\a}$ of the vector
multiplet instead of having bare superfield V terms.  Finally, in section 4 using Noether's procedure we
construct the superspace cubic interaction Lagrangian and find the coupling constant corrections to the free
gauge transformations.
%%%%%%%%%%%%%%%%%%%%%%%%%%%%
%%%%%%%%%%%%%%%%%%%%%%%%%%%%
\section{First order formulation}\label{sec:fof}
%%%%%%%%%%%%%%%%%%%%%%%%%%%%
%%%%%%%%%%%%%%%%%%%%%%%%%%%%
\subsection{Review of first order formulation of free higher spin fields}
\label{subsec:fofhs}
The power of first order formalism has been demonstrated repeatedly in the case of gravity, supergravity
and higher spins. Some of its characteristic features are (i) the independence between the vielbein and the
spin connection, (ii) the algebraic equation of motion for the spin connection, which can be solved in terms
of the vielbein field (and matter or gauge fields when present) and its derivative (iii)
the (algebraic) local Lorentz symmetry of the vielbein that reduces its degrees of freedom to match that of
the metric.

For higher spin fields a frame-like formalism has been developed by Vasiliev
\cite{Vasiliev:1980as,Vasiliev:1986td}. The key observation is that starting from the metric-like formulation using fully symmetric tensors
$h_{m_1m_2...m_s}$ $(~{\tiny\Yvcentermath1\overbrace{\yng(5)}^{\scriptsize s}}~)$ with appropriate gauge
transformations $\d h_{m_1m_2...m_s}=\pa_{(m_1}\xi_{m_2...m_s)}$ one can relax the symmetry for one of the
indices and define a not fully symmetric tensor field $e_{nm_1...m_{s-1}}$
(~${\tiny\Yvcentermath0\yng(1)\otimes\overbrace{\yng(5)}^{s-1}}$~) with gauge transformation $\d
e_{nm_1...m_{s-1}}=\pa_n\xi_{m_1...m_{s-1}}$.
The relaxed gauge transformation allows the construction of a first order, gauge invariant quantity
\begin{equation}
    I^{kl}{}_{m_1...m_{s-1}}\sim\epsilon^{pnkl}\pa_p e_{nm_1...m_{s-1}}
\end{equation}
which can be used to write a trivially invariant action $S[I]$. However, in order to maintain the same number of degrees of
freedom one must introduce an additional algebraic local symmetry $\d e_{nm_1...m_{s-1}}=a_{nm_1...m_{s-1}}$,
where $a_{nm_1...m_{s-1}}$ is an $(s-1,1)$ tableau (~${\tiny\Yvcentermath1\overbrace{\yng(4,1)}^{s-1}}$~).
Checking the invariance of $S[I]$ with respect to this local symmetry becomes more involved.
This process is simplified by realizing that $I^{kl}{}_{m_1...m_{s-1}}$ satisfies the following identity \begin{equation} \pa_k
I^{kl}{}_{m_1...m_{s-1}}=0 \end{equation} which can be promoted to a Bianchi identity that enforces the
$a$-transformation. This is achieved by
introducing
an auxiliary field, the connection, $\omega_{rnm_1...m_{s-1}}$
(~${\tiny\Yvcentermath1\yng(1)\otimes\overbrace{\yng(4,1)}^{s-1}}$~) with a transformation law
$\d \omega_{rnm_1...m_{s-1}}=\pa_r a_{nm_1...m_{s-1}}$.
In this description the action takes symbolically the form
\begin{equation}
    S[\omega,I]\sim\int~\omega^2 +\omega I~.\label{fofhs-action}
\end{equation}
This action provides a first order formulation of the usual Fronsdal description
\cite{Fronsdal:1978rb,Fang:1978wz} and by integrating out the connection
\footnote{For $s>2$ the action has an additional symmetry
$\d \omega_{rnm_1...m_{s-1}}=\sigma_{rnm_1...m_{s-1}}$, where $\sigma_{rnm_1...m_{s-1}}$ parameter is a
(s-1,2) tableau (~${\tiny\Yvcentermath1\overbrace{\yng(4,2)}^{s-1}}$~). This algebraic symmetry can be used to
introduce a tower of `extra' fields \cite{Vasiliev:1986td} which decouple at the free theory limit in order to
get second order equations of motion and can be understood as a realization of the de Wit-Freedman
connections.} one recovers the action $S[I]$.
%%%%%%%%%%%%%%%%%%%%%%%%%%%%%%%%%%%%%%%%%
%%%%%%%%%%%%%%%%%%%%%%%%%%%%%%%%%%%%%%%%%
\subsection{Superspace first order formulation of vector supermultiplet}
\label{subsec:superfofV}
It would be useful to have a similar first order description in superspace where supersymmetry
is manifest. Of course, there is the example of supergravity in superspace  which utilizes
supervielbeins and superconnections~\cite{Wess:1977fn,Siegel:1978mj,Ogievetsky:1979bb} (for details see
\cite{Gates:1983nr,Buchbinder:1995uq,RauschdeTraubenberg:2020kol}). However
such objects carry many more degrees of freedom and do not describe irreducible representations of the
super-Poincare group. Various sets of constraints have to be imposed  which are solved in terms of
prepotential superfields. The linearization of this description for the case of non-minimal supergravity
will give \eqref{masslessS}. By doing all that (solving the constraints and linearizing the theory), the
notion of a first order formulation, as described above, has faded and we effectively converted back to a
metric-like description. Therefore, our aspiration to acquire a first order formulation of linearized
supergravity and other higher spin supermultiplets that can be used to construct cubic interactions is a well
motivated one. In order to find one, we follow the spirit of the discussion of section
\ref{subsec:fofhs} but we use the unconstrained prepotential superfields instead of the conventional
superframes.

It will be instructive to start our consideration with a simpler supersymmetric multiplet, the vector multiplet
$(\Ysf=1/2)$. The superspace description of the vector multiplet is given by a real, scalar superfield $V$.
Its dynamics are determined by the following superspace action
\begin{equation}
    S_{(\Ysf=1/2)}[V]=\int d^8z ~V\D^{\g}\Dd^2\D_{\g}V
\end{equation}
which is invariant under the gauge transformation
\begin{equation}
    \d V=\D^2\bar{L}+\Dd^2 L~.
\end{equation}

As discussed in \ref{subsec:fofhs} we have to relax a property of the superfield $V$ which in turn
will simplify the gauge transformation and help us define a lower order invariant quantity. In the case of
spin 2 that property was the symmetrization of the two spacetime indices. In this case, the superfield $V$
does not carry any indices, so there is no symmetrization to relax. The only property that $V$ carries that
can be modified is it's reality ($V=\bar{V}$). Therefore, we drop the reality condition of $V$ and instead we
introduce a complex scalar superfield
$\mathcal{V}$ with the gauge transformation
\begin{equation}
    \d_{L} \V=\D^2\bar{L} \label{dLV}~.
\end{equation}
Due to the algebra of the supersymmetric covariant derivatives, one can immediately write a trivially gauge
invariant quantity
\begin{equation}
    \mathcal{I}_{\a}=\D_{\a}\V~,~\d_{L}\I_{\a}=0 \label{IV}~.
\end{equation}
By complexifying  $V$ we introduced extra degrees of freedom which must be removed. Following the
philosophy of first order formulation, we introduce an additional algebraic local symmetry
\begin{equation}
    \d_{\eta}\V=i\eta~,~\eta=\bar{\eta}
\end{equation}
which removes the imaginary part of $\V$ and thus restoring the degrees of freedom. Under this additional
$\eta$-transformation $\I_{\a}$ changes as follows:
\begin{equation}
    \d_{\eta}\I_{\a}=i\D_{\a}\eta~.
\end{equation}
The action will be constructed in
terms of $\I_{\a}$, making it trivially invariant under $L$- transformations. For the $\eta$-invariance
we introduce an auxiliary
superfield $\varOmega_{\a}$ playing the role of the connection and we select $\vO_{\a}$'s $\eta$-transformations
such that the action is also $\eta$-invariant. Because of it's auxiliary nature, $\vO_{\a}$ must appear in the
action algebraically, hence one is tempted to write an action similar to (\ref{fofhs-action}). However, due to
the engineering dimensions of the superspace integration measure this is impossible. The solution is that one
is
forced to introduce two connection-like superfields $\W_{\a}$ and $\vO_{\a}$. This is a familiar characteristic
feature of supersymmetric theories were the fermionic auxiliary (super)fields must appear in pairs
\cite{Gates:2017hmb} because they carry half odd-integer engineering dimensions and the superspace
Lagrangian has dimension 2 in four dimensions. Now, with all that in mind, we can write the most
general ansatz for the action:
\begin{equation}
S[\W,\vO,\I]=\int d^8z~\Bigg[c~\W^{\a}\vO_{\a}+\W^{\a}\I_{\a}+\vO^{\a}\Big\{
f_1\Dd^2\I_{\a}+f_2\D_{\a}\Dd^{\ad}\bar{\I}_{\ad}+f_3\Dd^{\ad}\D_{\a}\bar{\I}_{\ad}\Big\}+c.c.\Bigg]\label{SV}
\end{equation}
where coefficients $f_1, f_2, f_3$ are to be determined\footnote{Notice that the term $\D^2\I_{\a}$ is not
present inside the curly bracket. This term is zero due to (\ref{IV}).}. The variation
of this action can be organized as follows:
\begin{IEEEeqnarray*}{rl}
    \d S[\W,\vO,\I]=\int d^8z~\Bigg[&\d\W^{\a}\I_{\a}\n\label{dSV}
    +\d\vO^{\a}\Big\{f_1\Dd^2\I_{\a}+f_2\D_{\a}\Dd^{\ad}\bar{\I}_{\ad}+f_3\Dd^{\ad}\D_{\a}\bar{\I}_{\ad}\Big\}
    +c.c.\\
    +&\W^{\a}\Big\{c~\d\vO_{\a}+\d\I_{\a}\Big\}+c.c.\\
    +&\vO^{\a}\Big\{c~\d\W_{\a}+f_1\Dd^2\d\I_{\a}+f_2\D_{\a}\Dd^{\ad}\d\bar{\I}_{\ad}
+f_3\Dd^{\ad}\D_{\a}\d\bar{\I}_{\ad}\Big\}+c.c.\Bigg]
\end{IEEEeqnarray*}
Both $\eta$ and $L$-transformations of $\vO_{\a}$ and $\W_{\a}$ are chosen such that the last two lines
in (\ref{dSV}) vanish
\begin{IEEEeqnarray*}{l}\n\label{dWvO}
    \d\vO_{\a}=-\frac{1}{c}~\d\I_{\a}~,~
    \d\W_{\a}=-\frac{1}{c}~\Big\{f_1\Dd^2\d\I_{\a}+f_2\D_{\a}\Dd^{\ad}\d\bar{\I}_{\ad}
    +f_3\Dd^{\ad}\D_{\a}\d\bar{\I}_{\ad}\Big\}~,\sn\\[5pt]
    \d_{L}\vO_{\a}=0~,
    ~\d_{\eta}\vO_{\a}=-\frac{i}{c}~\D_{\a}\eta~,\sn\\
    \d_{L}\W_{\a}=0~,~
    \d_{\eta}\W_{\a}=-\frac{i}{c}(f_1+f_3)~\Dd^2\D_{\a}\eta+\frac{i}{c}(2f_2-f_3)~\D_{\a}\Dd^2\eta\sn~.
\end{IEEEeqnarray*}
Substituting (\ref{dWvO}) back in (\ref{dSV}) gives
\begin{IEEEeqnarray*}{l}
    0=\d_{\eta} S[\W,\vO,\I]=\frac{2i}{c}\int d^8z ~ \eta~\Big\{(f_1+f_3)~\D^{\a}\Dd^2\I_{\a}
    +(f_3-2f_2)~\Dd^2\D^{\a}\I_{\a}\Big\}+c.c.\n
\end{IEEEeqnarray*}
and fixes the coefficients
\begin{equation}
    f_1=-2f_2~,~f_3=2f_2
\end{equation}
With this choice, the action becomes:
\begin{equation}
    S[\W,\vO,\I]=\int d^8z~\Bigg[c~\W^{\a}\vO_{\a}+\W^{\a}\I_{\a}+f_2~\vO^{\a}\Big\{
-2\Dd^2\I_{\a}+\D_{\a}\Dd^{\ad}\bar{\I}_{\ad}+2\Dd^{\ad}\D_{\a}\bar{\I}_{\ad}\Big\}
    +c.c.\Bigg]~.\label{fSV}
\end{equation}
The equations of motion of $\W_{\a}$ and $\vO_{\a}$ are algebraic
\begin{equation}
    c~\vO_{\a}+\I_{\a}=0~,~c~\W_{\a}+f_2\Big\{
-2\Dd^2\I_{\a}+\D_{\a}\Dd^{\ad}\bar{\I}_{\ad}+2\Dd^{\ad}\D_{\a}\bar{\I}_{\ad}\Big\}=0
\end{equation}
and thus we can integrate them out. The result is the following action
\begin{equation}\label{SVfofw/oC}
    S_{(\Ysf=1/2)}[\I_{\a}]=-\frac{2f_2}{c}\int
            d^8z~\Big\{\I^{\a}\D_{\a}\Dd^{\ad}\bar{\I}_{\ad}+2~\I^{\a}\Dd^{\ad}\D_{\a}\bar{\I}_{\ad}
                -\Big(\I^{\a}\Dd^2\I_{\a}+c.c.\Big)\Big\}~.
\end{equation}
Using the definition of $\I_{\a}$ one can see that this action is equal to $S_{(\Ysf=1/2)}[V]$.
The equation of motion for superfield $\V$ in this
language takes the form
\begin{equation}
    \E^{(\V)}=\D^{\a}\Dd^2\I_{\a}+\Dd^{\ad}\D^2\bar{\I}_{\ad}~=~\D^{\g}\Dd^2\D_{\g}(\V+\bar{\V})=\D^{\a}W_{\a}
\end{equation}
where $W_{\a}=\Dd^2\D_{\a}V$ is the gauge invariant superfield strength for the vector
multiplet and $V=\V+\bar{\V}$. This  equation of motion satisfies the following Bianchi identity
\begin{equation}
    \E^{(\V)}=\bar{\E}^{(\V)}
\end{equation}
which is responsible for the $\eta$-invariance of $S_{(\Ysf=1/2)}[\I_{\a}]$.
%%%%%%%%%%%%%%%%%%%%%%%%%%%%%%%%%%%%%%%%%%%%%%%%%%%%%
%%%%%%%%%%%%%%%%%%%%%%%%%%%%%%%%%%%%%%%%%%%%%%%%%%%%%
\subsection{Superspace first order formulation of linearized non-minimal supergravity}
\label{subsec:superfofSG}
The superspace description of the non-minimal superspin $\Ysf=3/2$ supermultiplet
is given in terms of an unconstrained, real, gauge superfield $H_{\a\ad}$ and an unconstrained fermionic
compensator superfield $\chi_{\a}$
(we follow \cite{Gates:2013ska,Gates:2013tka}). The action is
\begin{IEEEeqnarray*}{rl}
    S_{({\scriptsize\Ysf=3/2})}[H_{\a\ad},\chi_{\a}]=\int d^8z~&\Big\{H^{\a\ad}\D^{\g}\Dd^2\D_{\g}H_{\a\ad}
                                    +2~\chi^{\a}\D_{\a}\Dd^{\ad}\bar{\chi}_{\ad}\n\label{masslessS}\\
                             &~-2H^{\a\ad}\Dd_{\ad}\D^2\chi_{\a}-2~\chi^{\a}\D^2\chi_{\a}~+c.c.\Big\}
\end{IEEEeqnarray*}
and is invariant under the following transformations
\begin{equation}
\delta H_{\a\ad}=\D_{\a}\bar{L}_{\ad}-\Dd_{\ad}L_{\a}~,~
\delta\chi_{\a}=\Dd^2L_{\a}+\D^{\b}\Lambda_{\a\b}~,~\Lambda_{\a\b}=\Lambda_{\b\a}~\n\label{dhx}~.
\end{equation}
The gauge invariant superfield strength is
\begin{equation}
    W_{\a\b\g}\sim\Dd^2\D_{(\a}\pa_{\b}{}^{\gd}H_{\g)\gd}~.
\end{equation}

Similar to the discussion in \ref{subsec:superfofV}, superfield $H_{\a\ad}$
has no symmetrization between indices, hence the only possible modification that one can do
is to relax its reality condition. Therefore, we introduce
a complex superfield $\mathcal{H}_{\a\ad}$ with the same index structure as the original $H$-superfield,
equipped with the gauge transformation
\begin{equation}
    \d_{L}\H_{\a\ad}= \D_{\a}\bar{L}_{\ad} \label{dLH}
\end{equation}
and an additional algebraic local symmetry that will fix the increase in the degrees of freedom
\begin{equation}
    \d_{\eta}\H_{\a\ad}=i\eta_{\a\ad}~,~\eta_{\a\ad}=\bar{\eta}_{\a\ad}~. \label{dnH}
\end{equation}
Next, we use the $L$-transformation \eqref{dLH} to find an invariant quantity that can be used to make the
action manifestly invariant. If we try something similar to \eqref{IV} we get
$\I_{\b\a\ad}=\D_{\b}\H_{\a\ad}$, but it is obvious that this is no longer an $L$-invariant because
$\d\I_{\b\a\ad}=-C_{\b\a}\D^2\bar{L}_{\ad}$, where $C_{\b\a}$ is the antisymmetric spinorial
metric\footnote{We use \emph{Superspace} \cite{Gates:1983nr} conventions.}. This can be fixed by
introducing a compensating superfield with an appropriate $L$-transformation. We introduce a
fermionic compensator
$\X_{\a}$ with the transformation
\begin{equation}
    \d_{L}\X_{\a}=\Dd^2L_{\a}~,~\d_{\eta}\X_{\a}=0~.
\end{equation}
The introduction of $\X_{\a}$ allows us to combine it with
$\H_{\a\ad}$ and define the following $L$-invariant quantity
\begin{equation}
    \I_{\b\a\ad}=\D_{\b}\H_{\a\ad}+C_{\b\a}\overline{\X}_{\ad}~,~\d_{L}\I_{\b\a\ad}=0~.\label{IH}
\end{equation}
The difference between \eqref{IV} and \eqref{IH}
is qualitative and can be traced back to the different number of supersymmetric covariant derivatives
participate in their corresponding $L$-transformations \eqref{dLV} and \eqref{dLH}.
The $\eta$-transformation of $\I_{\b\a\ad}$ can be easily found to be
\begin{equation}
    \d_{\eta}\I_{\b\a\ad}=i\D_{\b}\eta_{\a\ad}~.\label{dnI}
\end{equation}

The action will be written in terms of $\I_{\b\a\ad}$ and it's derivatives, making it manifestly
$L$-invariant. For the $\eta$-symmetry we rely on the critical role of the auxiliary connection-like
superfields we have to introduce. As discussed in \ref{subsec:superfofV}, because these superfields are
fermionic in nature, must come in pairs (~$\W_{\b\a\ad},~\vO_{\b\a\ad}$).
The general ansatz for this action (up to redefinitions) takes the form
\begin{IEEEeqnarray*}{l}
    S[\W,\vO,\I]=\int d^8z~\Bigg\{
    \W^{\b\a\ad}\vO_{\b\a\ad}+\W^{\b\a\ad}~\I_{\b\a\ad}+\vO^{\b\a\ad}\J_{\b\a\ad}\Bigg\}+c.c.\n\label{SHfof}
\end{IEEEeqnarray*}
where
\begin{IEEEeqnarray*}{rl}
    \J_{\b\a\ad}=&f_1~\D^2\I_{\b\a\ad}+f_2~\Dd^2\I_{\b\a\ad}+f_3~\D_{\b}\Dd^{\bd}~\bar{\I}_{\a\bd\ad}
                    +f_4~\Dd^{\bd}\D_{\b}~\bar{\I}_{\a\bd\ad}\n\\
                 &+g_1~C_{\b\a}~\Dd^2\I_{\ad}+g_2~\D_{\b}\Dd_{\ad}~\bar{\I}_{\a}
                     +g_3~\Dd_{\ad}\D_{\b}~\bar{\I}_{\a}\\
                 &+d_1~C_{\b\a}~\D^{\g}\Dd_{\ad}~\bar{\I}_{\g}+d_2~C_{\b\a}~\Dd_{\ad}\D^{\g}~\bar{\I}_{\g}
\end{IEEEeqnarray*}
and $\I_{\ad}=C^{\b\a}~\I_{\b\a\ad}$.
The variation of this action will generate six terms that can be organized in the following way
\begin{IEEEeqnarray*}{rl}
    \d S[\W,\vO,\I]=\int
    d^8z~\Bigg\{&~\d\W^{\b\a\ad}~\I_{\b\a\ad}+\d\vO^{\b\a\ad}\J_{\b\a\ad}+c.c.\n\label{dSH}\\
                             &+\W^{\b\a\ad}\big\{\d\vO_{\b\a\ad}+\d\I_{\b\a\ad}\big\}+c.c.\\
                             &+\vO^{\b\a\ad}\big\{\d\W_{\b\a\ad}+\d\J_{\b\a\ad}\big\}+c.c.~~~\Bigg\}~.
\end{IEEEeqnarray*}
This is suggestive of the transformation laws that we should assign to $\W_{\b\a\ad}$ and $\vO_{\b\a\ad}$:
\begin{IEEEeqnarray*}{l}\n
    \d\vO_{\b\a\ad}=- \d\I_{\b\a\ad}~,~\d\W_{\b\a\ad}=- \d\J_{\b\a\ad}~,\sn\label{dWvOH}\\[5pt]
    \d_{L}\vO_{\b\a\ad}=0~,~\d_{\eta}\vO_{\b\a\ad}=-i \D_{\b}\eta_{\a\ad}~,\sn\\
    \d_{L}\W_{\b\a\ad}=0~,~\d_{\eta}\W_{\b\a\ad}=-i~(f_2+f_4)~\Dd^2\D_{\b}\eta_{\a\ad}
    +i~(2f_3-f_4-g_2)~\D_{\b}\Dd^2\eta_{\a\ad}\sn\\
    \hspace{46mm}+i~g_1~C_{\b\a}~\Dd^2\D^{\g}\eta_{\g\ad}-i~d_1~C_{\b\a}~\D^{\g}\Dd^2\eta_{\g\ad}\\
\hspace{46mm}+i~g_3~\Dd_{\ad}\D_{\b}\Dd^{\gd}\eta_{\a\gd}+i~d_2~C_{\b\a}~\Dd_{\ad}\D^{\g}\Dd^{\gd}\eta_{\g\gd}~.
\end{IEEEeqnarray*}
Substituting the above transformations to \eqref{dSH} we find that the $\eta$-variation of the action is
\begin{IEEEeqnarray*}{l}
    \d_{\eta}S[\W,\vO,\I]=\int d^8z~i~\eta^{\a\ad}\Bigg\{
        2(f_2+f_4)~\D^{\b}\Dd^2~\I_{\b\a\ad}-(4f_3-2f_4-g_2)~\Dd^2\D^{\b}~\I_{\b\a\ad}\n\\
        \hspace{49mm} -g_3~\Dd_{\ad}\D^{\b}\Dd^{\bd}~\I_{\b\a\bd}+(2g_1-g_3)~\D_{\a}\Dd^2~\I_{\ad}\\
    \hspace{49mm} +(-2d_1+2g_2-g_3)~\Dd^2\D_{\a}~\I_{\ad}+2d_2~\Dd_{\ad}\D_{\a}\Dd^{\bd}~\I_{\bd}\Bigg\}+c.c.
\end{IEEEeqnarray*}
Because the terms inside the bracket are independent (not related algebraically), the $\eta$-invariance
of the action is achieved by choosing
\begin{equation}
    f_2=-f_4,~g_2=4f_3-2f_4,~g_3=0,~g_1=0,~d_1=g_2,~d_2=0~.\label{C1}
\end{equation}
% fixing $\J_{\b\a\ad}$ to be
% \begin{IEEEeqnarray*}{l}
%     \J_{\b\a\ad}=&f_1~\D^2\I_{\b\a\ad}-f_4~\Dd^2\I_{\b\a\ad}+f_3~\D_{\b}\Dd^{\bd}~\bar{\I}_{\a\bd\ad}
%                     +f_4~\Dd^{\bd}\D_{\b}~\bar{\I}_{\a\bd\ad}\n\\
%                  &+2~(2f_3-f_4)~\D_{\b}\Dd_{\ad}~\bar{\I}_{\a}
%                  +2~(2f_3-f_4)~C_{\b\a}~\D^{\g}\Dd_{\ad}~\bar{\I}_{\g}
% \end{IEEEeqnarray*}

Unlike the vector multiplet case, this procedure did not fix all the parameters and there is still
plenty of freedom left parametrized by coefficients $f_1,~f_3,~f_4$. This is a signal that there is some
further structure to be explored. Interestingly, one can check that $\I_{\b\a\ad}$ satisfies the following
two identities:
\begin{IEEEeqnarray*}{l}\n\label{BI}
    \D_{(\g}~\I_{\b\a)\ad}=0~,\sn\label{BI1}\\
    \D_{(\b}\Dd^{\ad}\D_{\a)}~\I_{\ad}
    -2~\D_{(\b}\Dd^{\ad}\D^{\g}~\I_{|\g|\a)\ad}+3\D^2\Dd^{\ad}\I_{(\b\a)\ad}=0~.\sn\label{BI2}
\end{IEEEeqnarray*}
Interpreting these two identities as hints for additional symmetries, we promote them
to Bianchi identities which are responsible for the invariance of the action under these symmetries.
For example, \eqref{BI1} suggests
that in the $\W^{\b\a\ad}~\I_{\b\a\ad}$ term of the action \eqref{SHfof} we can shift
$ \W_{\b\a\ad}\to\W_{\b\a\ad}+\D^{\g}\Lambda_{\g\b\a\ad} $
% \begin{equation}
%     \W_{\b\a\ad}\to\W_{\b\a\ad}+\D^{\g}\Lambda_{\g\b\a\ad}
% \end{equation}
for some $\Lambda_{\g\b\a\ad}$ which is symmetric in the undotted indices. To explore this possibility
we introduce a new transformation
\begin{equation}
\d_{\L}\H=0~,~\d_{\L}\X_{\a}=\D^{\g}\L_{\g\a}~,~\L_{\a\b}=\L_{\b\a}~\Rightarrow~
\d_{\L}\I_{\b\a\ad}=-~C_{\b\a}~\Dd^{\gd}\bar{\L}_{\gd\ad}~.\label{dLI}
\end{equation}
Using rule \eqref{dWvOH} we find the appropriate $\L$-transformations for the two connection-like superfields
\begin{IEEEeqnarray*}{l}
    \d_{\L}\vO_{\b\a\ad}=C_{\b\a}~\Dd^{\gd}\bar{\L}_{\gd\ad}~,~
    \d_{\L}\W_{\b\a\ad}=f_1~C_{\b\a}~\D^2\Dd^{\gd}\L_{\gd\ad}+(f_3-2g_2)~\D_{\b}\Dd_{\ad}\D^{\g}\L_{\g\a}\n\\
    \hspace{58mm}+(2g_3-f_4)~\Dd_{\ad}\D^2\L_{\b\a}-2d_1~C_{\b\a}~\D^{\g}\Dd_{\ad}\D^{\r}\L_{\r\g}
\end{IEEEeqnarray*}
and the $\L$-variation of the action (using \eqref{dSH} and \eqref{BI2}) becomes
\begin{IEEEeqnarray*}{l}
    \d_{\L}S[\W,\vO,\I]=\int d^8z ~ \L^{\g\a}\Bigg\{-2f_1~\D_{\g}\Dd^2~\bar{\I}_{\a}
        -2(f_3-2g_2+4d_1)~\D_{\g}\Dd^{2}\D^{\b}~\I_{\b\a\ad}\n\\
    \hspace{53mm}   +(2f_4-2g_3+12d_1-3g_2)~\D^2\Dd^{\ad}~\I_{\g\a\ad}\Bigg\}+c.c.
\end{IEEEeqnarray*}
The $\L$-invariance of the action fixes the remaining free parameters such that
\begin{equation}
    f_1=0~,~f_3=2g_2-4d_1~,~f_4=g_3-6d_1+\frac{3}{2}g_2
\end{equation}
which together with \eqref{C1} fix all coefficients:
\begin{equation}
f_1=0~,~f_2=-f_4~,~f_3=\frac{4}{9}f_4~,~~~
g_1=0~,~g_2=-\frac{2}{9}f_4~,~g_3=0~,~~~
d_1=-\frac{2}{9}f_4~,~ d_2=0~.
\end{equation}
For these coefficients the $\L$-transformation of $\W_{\b\a\ad}$ takes the form
\begin{equation}
    \d_{\L}\W_{\b\a\ad}=\frac{4}{3~3!}~\D^{\g}\Dd_{\ad}\D_{(\g}~\L_{\b\a)}
    +\frac{4}{3}~\D^2\Dd_{\ad}\L_{\b\a}+\frac{1}{3}~\Dd_{\ad}\D^2~\L_{\b\a}
\end{equation}
and
the superspace action of the linearized $4D,~\N=1$ non-minimal supergravity
supermultiplet in this first order formulation is
\begin{IEEEeqnarray*}{l}\n\label{Hfof}
    S[\W,\vO,\I]=\int d^8z~\Bigg\{~~~
    \W^{\b\a\ad}\vO_{\b\a\ad}+\W^{\b\a\ad}~\I_{\b\a\ad}\\
    \hspace{40mm}-c~\vO^{\b\a\ad}
\Bigg[\Dd^2~\I_{\b\ad}-\frac{4}{9}~\D_{\b}\Dd^{\bd}~\bar{\I}_{\a\bd\ad}-\Dd^{\bd}\D_{\b}~\bar{\I}_{\a\bd\ad}\\
\hspace{63mm}+\frac{2}{9}~\D_{\b}\Dd_{\ad}~\bar{\I}_{\a}+\frac{2}{9}~C_{\b\a}~\D^{\g}\Dd_{\ad}~\bar{\I}_{\g}
~~~\Bigg]~ \Bigg\} +c.c.
\end{IEEEeqnarray*}
One can do one more step and integrate out the auxiliary connection-like superfields.
The equations of motion for $\W_{\b\a\ad}$
and $\vO_{\b\a\ad}$ are:
\begin{IEEEeqnarray*}{l}\n
    \vO_{\b\a\ad}+\I_{\b\a\ad}=0~,~\sn\\[5pt]
   -\frac{1}{c}~\W_{\b\a\ad}+
\Dd^2~\I_{\b\ad}-\frac{4}{9}~\D_{\b}\Dd^{\bd}~\bar{\I}_{\a\bd\ad}-\Dd^{\bd}\D_{\b}~\bar{\I}_{\a\bd\ad}
+\frac{2}{9}~\D_{\b}\Dd_{\ad}~\bar{\I}_{\a}+\frac{2}{9}~C_{\b\a}~\D^{\g}\Dd_{\ad}~\bar{\I}_{\g}=0~~~~~.\sn
\end{IEEEeqnarray*}
Substituting them in \eqref{Hfof} we find the following action
\begin{IEEEeqnarray*}{l}\n\label{SIH}
    S_{(\Ysf=3/2)}[\I_{\b\a\ad}]=c\int
    d^8z~\Bigg\{\I^{\b\a\ad}~\Dd^2~\I_{\b\a\ad}+\frac{2}{9}~\I^{\b\a\ad}~\D_{\b}\Dd_{\ad}~\bar{\I}_{\a}~~+c.c.\\
        \hspace{47mm}        -\frac{8}{9}~\I^{\b\a\ad}~\D_{\b}\Dd^{\bd}~\bar{\I}_{\a\bd\ad}
        -2~\I^{\b\a\ad}~\Dd^{\bd}\D_{\b}~\bar{\I}_{\a\bd\ad}
    +\frac{4}{9}~\I^{\ad}~\D^{\a}\Dd_{\ad}~\bar{\I}_{\a}~~\Bigg\}~.
\end{IEEEeqnarray*} It is straightforward to show that this action is equal to $S_{(\Ysf=3/2)}[H_{\a\ad},
\chi_{\a}]$ with the identification $H_{\a\ad}=\H_{\a\ad}+\bar{\H}_{\a\ad}$ and $\chi_{\a}=\X_{\a}$.
The equations of motion for $\H_{\a\ad}$ and $\X_{\a}$ as derived from the above action are
\begin{IEEEeqnarray*}{l}\n
    \E^{(\H)}_{\a\ad}=\D^{\b}\Dd^2~\I_{\b\a\ad}+\Dd^{\bd}\D^2~\overline{\I}_{\a\bd\ad}~,\sn\\
    \overline{\E}^{(\X)}_{\ad}=\Dd^2~\I_{\ad}+\frac{1}{3}~\D^{\a}\Dd_{\ad}~\overline{\I}_{\a}
    +\frac{2}{3}~\D^{\a}\Dd^{\bd}~\overline{\I}_{\a\bd\ad}+\Dd^{\bd}\D^{\a}~\overline{\I}_{\a\bd\ad}~.\sn
\end{IEEEeqnarray*}
These equations of motion satisfy the following Bianchi identities
\begin{IEEEeqnarray*}{ll}\n
    \D^{\a}~\E^{(\H)}_{\a\ad}=\D^2~\overline{\E}^{(\X)}_{\ad}~~~&\Big[L-\text{invariance}\Big]\sn\\
    \E^{(\H)}_{\a\ad}=\overline{\E}^{(\H)}_{\a\ad}  &\Big[\eta-\text{invariance}\Big]\sn\\
    \Dd_{(\bd}~\overline{\E}^{(\X)}_{\ad)}=0  &\Big[\L-\text{invariance}\Big]\sn
\end{IEEEeqnarray*}
A practical advantage of action \eqref{SIH} is that
it packages both $\H_{\a\ad}$ and $\X_{\a}$ superfields into one object $\I_{\b\a\ad}$. This is a very
attractive feature if we want to construct interactions.

It is useful to compare the structures that appear in the first order formulation as described
above, with the corresponding structures in the conventional superspace description of
supergravity. In both cases we have unconstrained prepotential superfields, connections and an enhanced set of
symmetries. However, in the first order description some of these features are motivated for different reasons.

In conventional supergravity formulation, one starts with the algebra of covariant derivatives constructed out
of supervielbeins and superconnections. Then various constraints are imposed on supertorsion in order to
minimize the  number of
independent superfields.  These constraints are solved by a set of
unconstrained (prepotential) superfields, which determine the superconnection and supervielbein. Moreover, the
prepotentials carry additional symmetries that preserve the constraints. Besides the general super-coordinates
transformation and the local super-Lorentz rotations an additional symmetry appears ($\L$ symmetry). The
algebraic terms of these symmetries can be used to remove unnecessary prepotentials, for example the real part
of the complex vector prepotential.

From the viewpoint of the linearized theory \eqref{masslessS}, the same prepotentials emerge as the
appropriate variables for the Lagrangian description of the corresponding irreducible representation, as
determined by the diagonalization of the Casimir operators of supersymmetry algebra. In this configuration
there is no longer an algebraic local symmetry and the rest of the surviving symmetries are linearized and
take the form of gauge transformations \eqref{dhx}.

In the first order formulation \eqref{Hfof} the notion of an algebraic local symmetry is restored by
complexifying the vector prepotential and introduce the local $\eta$-transformation \eqref{dnH} in
order to preserve the degrees of freedom of the theory. In this case, the complexification is motivated by the
`first order' philosophy which probes for invariant quantities with less derivatives than the equations of
motion. Moreover, the connection-like superfields are
introduced at the level of the action as auxiliary superfields that help us achieve $\eta$-invariance and
are determined by their algebraic equations of motion. Interestingly, this formulation can be
extended to higher spin supermultiplets, in contrast to the covariant derivative approach
of supergravity in which this extension is not obvious. We will not explore this direction in this work, but it will be investigated
in a follow up paper.
%%%%%%%%%%%%%%%%%%%%%%%%%%%%%%%%%%%%%%%%%%%%%%%%%%%%
%%%%%%%%%%%%%%%%%%%%%%%%%%%%%%%%%%%%%%%%%%%%%%%%%%%%
\section{Trivial symmetries and Noether's procedure}
\label{sec:TrSym}
\subsection{Review of Trivial symmetries with examples in supersymmetric theories}
Let us consider a theory that includes multiple copies of the same kind of fields. Then, we can streamline the
notation and combine all these fields into one object $\phi^{I}$ by introducing an index $I=1,2,...$ that
counts them. For simplicity we suppress all other types of indices that these fields may carry\footnote{
    Index $I$ as introduced here may appear to be an internal index. However all arguments
    can be restated even if $I$ is replaced by spinorial or spacetime indices. They
also count the number of components of the fields.}. Now let us consider the following
transformation
\begin{equation}
    \d_{\l}\phi^{I}=\l~C^{IJ}~\frac{\d S}{\d\phi^{J}}\label{pts}
\end{equation}
where $\l$ is the parameter that controls the transformation, $C^{IJ}$ is an appropriate collection of
coefficients and $\frac{\d S}{\d\phi^{J}}$ is the equation of motion for field  $\phi^{J}$. The
variation of the action of this theory $S[\phi^{I}]$ under the above $\l$-transformation (with appropriate boundary
conditions) is
\begin{equation}
    \d_{\l}S=\int \l~C^{IJ}~ \frac{\d S}{\d\phi^{J}}~\frac{\d S}{\d\phi^{I}}~.
\end{equation}
By choosing appropriately the symmetry property of $C^{IJ}$ this variation vanishes. For example, if
$\phi^{I}$ are commuting [anticommuting] fields then by choosing $C^{IJ}$ to be antisymmetric
($C^{IJ}=-C^{JI}$) [symmetric ($C^{IJ}=C^{JI}$)] we get
$\d_{\l}S=0$. Hence the $\l$-transformation is a symmetry of the theory.

These transformations are called \emph{trivial symmetries} and form an infinite class of obvious invariants that
every theory has. Trivial symmetries do not reduce the number of physical degrees of freedom, for this reason one is
usually not interested in such transformations and they are not even included in the set of
symmetries. However, it can be fruitful to be aware of these trivial symmetries when constructing
interactions. In some cases, the parameter $\l$ can be proportional to the coupling constant of
the interaction and hence the combination of the trivial symmetries with Noether's procedure can reveal
properties about the structure of the interaction vertex. A successful application of this approach can be
found in \cite{Zinoviev:2008jz}.

Furthermore, because the commutator of infinitesimal symmetries is also an infinitesimal symmetry of the
action, trivial symmetries may appear in the right hand side of the commutators of generators of
symmetries.
% \footnote{$ \big[\d_{S_1}~,~\d_{S_2}\big]=\text{linear combination of symmetries including trivial
% symmetries}$.}.
Famously, this is the case in some formulations of supersymmetric theories where one can find deformations of
the usual supersymmetry algebra by terms proportional to the equations of motion. This phenomenon is sometimes
called as `on-shell SUSY closure'. In all such cases, these deformation terms are precisely trivial
symmetries that emerge in the right hand side of the commutator.
The identifying characteristic is not only that they are proportional to the equations of motion
but also crucially the proportionality coefficients have the correct symmetry property to make them trivial
symmetries
\footnote{For example the commutator of two supersymmetry transformations of the chiral multiplet fermion
after integrating out the auxiliary fields takes the form
$\big\{\e^{a}\mathcal{Q}_{a}~,~\eta^{b}\mathcal{Q}_{b}\big\}\psi^{c}\sim i(\e\g^{m}\eta)\pa_{m}\psi^{c}
+\underbrace{(\e\g^{n}\eta)}_{\l}~\underbrace{(\g_n)^{cd}}_{C^{cd}=C^{dc}}~\frac{\d S}{\d\psi^{d}}$}.
In fact, from the view point of supersymmetric theories formulated in a non manifestly manner (such as the
component formulation) one way to justify the existence of auxiliary fields is the removal of these
trivial symmetry
deformations of the algebra such that we get honest representations of supersymmetry algebra.

A less known application of trivial symmetries in supersymmetric theories has to do with a special class of
theories with auxiliary fermions in their off-shell spectrum.
Non-minimal supergravity is such a theory, as well as its higher spin cousins.
As mentioned previously, the auxiliary fermions come
in pairs $(\b,\r)$ with different engineering dimensions and the action includes a term
$\b^{\a}\r_{\a}$. In that case, there is a trivial symmetry between
the auxiliary fermions $\b,\r$ and the dynamical fermions $\psi$
with the transformation parameter being dimensionless\footnote{
    The simplest example is the complex linear supermultiplet. The transformation
    $\d\b=c~\g^{m}\pa_{m}\psi$, $\d\psi=-c~\r$, $\d\r=0$ is a trivial symmetry of the theory. Notice that the
equation of motion of the auxiliary fermion $\b$ is $\E^{(\b)}=\r$.}. For these theories, the trivial
symmetries can be used to simplify the supersymmetry transformation of the dynamical fermion $\d_{Q}\psi$
(by removing all terms proportional to $\d_{Q}\r$).
\subsection{Application to electromagnetic interactions}
\label{subsec:TrSymEI}
A characteristic feature of electromagnetic interactions of various fields is that because of the underlying
$U(1)$ (locally isomorphic to SO(2)) symmetry, these fields come in doublets. Therefore trivial symmetries like
\eqref{pts} emerge.
In this case, we have a doublet of  $\Ysf=3/2$ supermultiplets labeled by an internal
index $i$ which takes two values. Hence, following the discussion above there are trivial
symmetries that preserve the sum of the two free actions. An example of such a trivial symmetry is
\begin{equation}
    \d H^{i}_{\a\ad}=\frac{\l}{M^2}~\epsilon^{ij}~\E^{j~(H)}_{\a\ad}\label{ts}
\end{equation}
where $\E^{j(H)}_{\a\ad}$ is the equation of motion of $H^{j}_{\a\ad}$\footnote{We can also write this
trivial symmetry in terms of the complexified superfield $\H_{\a\ad}$ introduced in \ref{subsec:superfofSG}.},
$M$ is an appropriate mass scale
required by engineering dimensions, $\l$ is a dimensionless, arbitrary, real
superfield and $\e^{ij}$ is the unique
(up to an overall factor) two dimensional, rank 2, antisymmetric tensor. Although we know that this
transformation is nothing more than a trivial symmetry, because it introduces a dimensionful parameter and is
linear in superfield $H_{\a\ad}$ it can
be interpreted as a trivial\footnote{Trivial because it does not correspond to a non-trivial
interaction vertex}, first order (in the coupling constant) correction to the free gauge transformations. In
this case, the coupling constant is  $1/M^2$ which we know from Metsaev's classification \cite{Metsaev:2005ar}
is the correct coupling constant for the cubic interaction between two spin 2 gauge fields and a spin 1 gauge
field involving three spacetime derivatives. This is also consistent with the non-supersymmetric 2-2-1 vertex
\cite{Boulanger:2008tg,Zinoviev:2008jz} which will be included in the supersymmetric $3/2-3/2-1/2$ vertex that
we are constructing.

In this interpretation of \eqref{ts}, the dimensionless real scalar superfield $\l$ is a function of the free
gauge transformation parameters. For the vector multiplet the gauge parameter is an
arbitrary dimensionless
 chiral superfield $\Phi$ $(\d V=\Phi+\bar{\Phi}~,\Phi=\Dd^2 L)$ hence we can write
\begin{equation}
    \d^{(\Phi)}_1
    H^{i}_{\a\ad}=\frac{c}{M^2}\big(\Phi+\bar{\Phi}\big)~\e^{ij}~\E^{(H)}_{j~\a\ad}\label{d1Phi}~.
\end{equation}
Using Noether's procedure of constructing consistent interactions perturbatively we must satisfy the
following gauge consistency condition for cubic vertices
\begin{equation}
    \d^{(\xi)}_0 S_{1}+\d^{(\xi)}_1 S_0=0
\end{equation}
where $\d^{(\xi)}_{0},~\d^{(\xi)}_{1}$ are the zeroth and first order variations with respect to
transformations controlled by parameter $\xi$. $S_{0},~S_{1}$ are the free and first order correction actions
respectively. Applying this for $\xi=\Phi$ and using the fact that \eqref{d1Phi} is a trivial symmetry
($\d^{(\Phi)}_{1}S_{0}=0$) we conclude that the cubic interaction must be trivially invariant under the free
gauge transformation of the vector multiplet
\begin{equation}
    \d^{(\Phi)}_{0}S_{1}=0~.
\end{equation}
The only way this can be true is if the interaction vertex depends on the gauge invariant superfield strength
$W_{\a}=\Dd^2\D_{\a}V$ instead of the bare vector superfield $V$.
%%%%%%%%%%%%%%%%%%%%%%%%%%%%%%%%%%%%%%%%%%%%%%%%%%%%%%%%%%%%%%%%
%%%%%%%%%%%%%%%%%%%%%%%%%%%%%%%%%%%%%%%%%%%%%%%%%%%%%%%%%%%%%%%%
\section{Electromagnetic interaction of non-minimal linearized supergravity}
\label{sec:ElectroSugra}
\subsection{Preparations}
Our aim is to find a consistent cubic interaction between the vector multiplet and a doublet of
linearized non-minimal
supergravity multiplets. For the description of the vector multiplet and the two supergravity multiplets
we will use the formulations developed in \ref{subsec:superfofV} and \ref{subsec:superfofSG}.
Our starting action is the sum of free actions \eqref{SIH} and \eqref{SVfofw/oC}
\begin{equation}
    S_{0}=\sum_{i=1}^{2}~S_{(\Ysf=3/2)}[\I^{i}_{\b\a\ad}]~+~c~S_{(\Ysf=1/2)}[\I_{\a}]
\end{equation}
where $c$ is a fixed relative coefficient between the two free actions which ensures that (up to an overall
sign) both actions have positive kinetic energy. We search for $S_1$, the first order correction to $S_0$
in the coupling constant expansion. We know that it will be quadratic in the supergravity superfields, linear
in the vector multiplet, the coupling constant is $1/M^{2}$ and it  depends on the superfield strength
$W_{\a}=\Dd^2\D_{\a}V$ \big($S_1=S_1[\I^{i}_{\b\a\ad},W_{\a}]$\big). Also, a
quick dimensional analysis argument shows that it must involve $3$  supersymmetric covariant derivatives
distributed among $\I^{i}_{\b\a\ad}$ and $W_{\a}$. Finally, it must satisfy the consistency conditions
\begin{equation}
    \d^{(\eta)}_{1}S_{0}+\d^{(\eta)}_{0}S_{1}=0~,~
    \d^{(\L)}_{1}S_{0}+\d^{(\L)}_{0}S_{1}=0\label{Nc}
\end{equation}
where $\d^{(\eta)}_{0},~\d^{(\L)}_{0}$ are the variations according to transformations
\eqref{dnI} and \eqref{dLI} respectively and $\d^{(\eta)}_{1},~\d^{(\L)}_{1}$ are appropriate first order
corrections to the corresponding free transformations. Equations \eqref{Nc} must be solved
simultaneously for $S_1,\d_{1}^{(\eta)},\d_{1}^{(\L)}$.

A conventional approach to solve these equations is to make an ansatz for $S_1$ and calculate it's variation
under free gauge transformations with the assumption that the on-shell equations of motion for the free theory
hold. On-shell, Noether's constraints simplify to $\d_{0}S_1\big|_{\d S_{0}=0}=0$ which can be solved
by fixing the coefficients in our ansatz. Having found a non-trivial $S_1$ then the calculation can be repeated
off-shell in order to find $\d_{1}$. In our case, just writing a non-trivial ansatz with all possible index
contractions and various derivative distributions generates a very large number of terms that can not be
controlled easily. For this reason, we will start in the opposite end and consider various ansatz for the
first order corrections of the transformation laws. There are not many terms that one can write  for $\d_1$
and even if we do not include everything, just a few terms are enough to get us started and generate
everything else in the process. This procedure will give us clues regarding $S_1$.

For the corrections to the $\eta$ transformation of superfield $\H^{i}_{\a\ad}$ we can write
\begin{equation}
    \d^{(\eta)}_{1}\H^{i}_{\a\ad}=i\frac{d}{M^2}~\e^{ij}~\eta_{j}^{\b}{}_{\ad}~\D_{(\b}W_{\a)}+...
\end{equation}
where the dots correspond to additional terms that we can write. Let us consider the effect of this deformation
in the transformation law by varying only $\H^{i}_{\a\ad}$ in $S_{0}$. After some lengthy calculations we find
\begin{IEEEeqnarray*}{l}
    \d^{(\eta)}_{1}S_{0}=-\frac{id}{M^2}\int d^8z~\e^{ij}\Bigg\{
        \D^{\b}\eta^{j}_{(\b}{}^{\ad}~W^{\a}~\E^{i~(\H)}_{\a)\ad}
        -\eta^{j}_{(\b}{}^{\ad}~\D^{\a}\Dd^{\gd}\D^{\b}\overline{\I}^{i}_{\a)\gd\ad}~\E^{(\V)}\n\label{d1nS0}\\
        \hspace{49mm}+\D_{\r}\eta^{j}_{(\b}{}^{\ad}~W^{\r}~\Big[
            \D^{\a}\Dd^{\gd}\D^{\b}\overline{\I}^{i}_{\a)\gd\ad}
    -\D^{\a}\Dd^2\I^{i\b}{}_{\a)\ad}\Big]\Bigg\}+c.c.
\end{IEEEeqnarray*}
The first two terms are proportional to the equations of motion of superfields $\H^{i}_{\a\ad}$ and $\V$, hence
they can be compensated by adding appropriate corrections in their corresponding transformation laws. The rest
of the terms are non-trivial because they can only be
canceled by $\d_{0}^{(\eta)}S_1$. This is a suggestion to consider the following ansatz for $S_{1}$
\begin{equation}
    S_1^{(a)}=\frac{1}{M^2}\int d^8z~\e^{ij}~\I^{j}_{\r(\b}{}^{\ad}~W^{\r}\Big(
    \D^{\a}\Dd^{\gd}\D^{\b}\overline{\I}^{i}_{\a)\gd\ad} -\D^{\a}\Dd^2\I^{i\b}{}_{\a)\ad}\Big)+c.c.
\end{equation}
Using \eqref{dnI}, one can show that the zeroth order variation of the terms inside the
parenthesis vanishes, hence $\d^{(\eta)}_0S^{(a)}_{1}$ matches precisely the terms in the second line of
\eqref{d1nS0}.  Therefore an appropriate choice of parameter $d$ will solve the $\eta$-Noether consistency
condition in \eqref{Nc}.

For the $\L$-Noether condition, we calculate $\d^{\L}_{0}S^{(a)}_{1}$. Again after some algebra we find
\begin{IEEEeqnarray*}{l}
    \d^{(\L)}_{0}S^{(a)}_{1}=\frac{1}{M^2}\int d^8z~\e^{ij}\Bigg\{
        \Dd_{\rd}\bar{\L}^{j~\rd\ad}~W^{\b}~\E^{i~(\H)}_{\b\ad}
        +2~\D_{\b}\Big(\Dd_{\rd}\bar{\L}^{j~\rd\ad}~W^{\b}\Big)~\overline{\E}^{i~(\X)}_{\ad}\n\label{d1LS1a}\\
        \hspace{47mm}-\frac{1}{3}~\Dd_{\rd}\bar{\L}^{j\rd\ad}~W^{\a}\Big[
    2~\D^2\Dd_{\ad}\overline{\I}^{i}_{\a}+\D^2\Dd^{\gd}\overline{\I}^{i}_{\a\gd\ad}\Big]\\
\hspace{47mm}+\I^{j\r\b\ad}~W_{\r}~\D^2\Dd_{\ad}\D^{\a}\L^{i}_{\a\b}\Bigg\}+c.c.
\end{IEEEeqnarray*}
Observe that both terms in the fist line are proportional to the equation of motion of $\H^{i}_{\a\ad}$ and
$\X^{i}_{\a}$ superfields, hence these terms can be eliminated by appropriate deformations in there
$\L$-transformations. However, the rest of the terms are not of this kind, hence $S^{(a)}_{1}$ is not enough
to solve \eqref{Nc}, we have to consider additional contributions. Equation \eqref{d1LS1a} provides a hint for such an
additional contribution
\begin{equation}
    S_{1}^{(b)}=\frac{g}{M^2}\int
    d^8z~\e^{ij}~\I^{j~}_{\r}{}^{\g\ad}~\W^{\r}~\D^2\Big(
    2~\Dd_{\ad}\overline{\I}^{i}_{\g}+\Dd^{\gd}\overline{\I}^{i}_{\g\gd\ad}\Big)+c.c.
\end{equation}
It is straightforward to check that under transformations \eqref{dLI}, $\d^{(\L)}_{0}S^{(b)}_{1}$ matches
precisely the last two lines of \eqref{d1LS1a} up to an overall constant which can be adjusted appropriately
and thus solving the $\L$-Noether constraint.

However, by adding $S^{(b)}_{1}$ we have ruined the $\eta$-invariance, thus one must repeat the above
calculations for the updated $S_1$ ansatz
\begin{equation}
    S_1=S^{(a)}_1+S^{(b)}_1~.
\end{equation}
In general, we may have to keep adding terms $S_1^{(c)},~S^{(d)}_1,...$ in order to counteract
the effect of previous term. However in this case, terms $S^{(a)}_{1}$ and $S^{(b)}_{1}$ are enough to
satisfy \eqref{Nc} and to determine the deformations of the gauge transformations.
\subsection{Cubic interaction $3/2-3/2-1/2$}
Consider the ansatz
\begin{IEEEeqnarray*}{l}
    S_1=
    \frac{1}{M^2}\int d^8z~\e^{ij}\Bigg\{~\I^{j}_{\r(\b}{}^{\ad}~W^{\r}~\D^{\a}\Big(
        \Dd^{\gd}\D^{\b}\overline{\I}^{i}_{\a)\gd\ad} -\Dd^2\I^{i~\b}{}_{\a)\ad}\Big)\n\\
        \hspace{39mm}+g~\I^{j~}_{\r}{}^{\g\ad}~\W^{\r}~\D^2\Big(
    2~\Dd_{\ad}\overline{\I}^{i}_{\g}+\Dd^{\gd}\overline{\I}^{i}_{\g\gd\ad}\Big)~~\Bigg\}+c.c.
\end{IEEEeqnarray*}
where an overall constant can be absorbed in the definition of the coupling constant.
We calculate it's $\L$ and $\eta$-variations using \eqref{dLI} and \eqref{dnI} respectively:
\begin{IEEEeqnarray*}{l}
    \d^{(\L)}_{0}S_{1}=\frac{1}{M^2}\int
    d^8z~\e^{ij}\Bigg\{~~2~\D_{\r}\Big(\Dd_{\sd}\bar{\L}^{j\sd\ad}~W^{\r}\Big)~\overline{\E}^{i~(\X)}_{\ad}\n
        \label{dLS1}\\
        \hspace{46mm}+~\Dd_{\sd}\bar{\L}^{j\sd\ad}~W^{\r}~\E^{i~(\H)}_{\r\ad}\\
        \hspace{46mm}+(3g-1)~\I^{j~}_{\r}{}^{\b\ad}~W^{\r}~\D^{\a}\Dd_{\ad}\D^2\L^{i}_{\b\a}\\
        \hspace{46mm}-(g-\frac{1}{3})~\Dd_{\sd}\bar{\L}^{j\sd\ad}~W^{\r}~\D_{\r}\Big(
    2~\D^{\a}\Dd_{\ad}\overline{\I}^{i}_{\a}+\D^{\a}\Dd^{\gd}\overline{\I}^{i}_{\a\gd\ad}\Big)\Bigg\}+c.c.
\end{IEEEeqnarray*}
\begin{IEEEeqnarray*}{l}
    \d^{(\eta)}_{0}S_1=\frac{i}{M^2}\int
    d^8z~\e^{ij}\Bigg\{\Big[~~\eta^{j}_{(\b}{}^{\ad}\Big(
        \D^a\Dd^{\gd}\D^{\b}\overline{\I}^{i}_{\a)\gd\ad}-\D^{\a}\Dd^2\I^{i\b}{}_{\a)\ad}\Big)\n\\
\hspace{55mm}+g~\eta^{j\g\ad}~\D^2\Big(2~\Dd_{\ad}\overline{\I}^{i}_{\g}+\Dd^{\gd}\overline{\I}^{i}_{\g\gd\ad}
\Big)\Big]
~\E^{(\V)}\\
\hspace{51mm}-~\D^{\b}\Big(\eta^{j}_{(\b}{}^{\ad}~W^{\a}\Big)~\E^{i~(\H)}_{\a)\ad}~\Bigg\}+c.c.
\end{IEEEeqnarray*}
Similarly, the variation of $S_{0}$ under the first order transformations
\begin{IEEEeqnarray*}{l}
    \d_{1}S_{0}=\int d^8z~\Bigg\{~-2~\d_{1}\H^{i~\a\ad}~\E^{i~(\H)}_{\a\ad}
        +2~\d_1\bar{\X}^{i~\ad}~\overline{\E}^{i~(\X)}_{\ad}
    -2c~\d_1\V~\E^{(\V)}~\Bigg\}+c.c.
\end{IEEEeqnarray*}
Therefore, \eqref{Nc} are solved by choosing $g=\frac{1}{3}$ and
\begin{IEEEeqnarray*}{l}\n
    \d_{1}^{(\eta)}\H^{i\a\ad}=-\frac{i}{2M^2}~\e^{ij}~\D^{(\b}\Big(\eta^{j}_{\b}{}^{\ad}~W^{\a)}\Big)~,~~~
    \d_{1}^{(\L)}\H^{i\a\ad}=\frac{1}{2M^2}~\e^{ij}~\Dd_{\sd}\bar{\L}^{j\sd\ad}~W^{\a}~,\sn\\[6pt]
    \d_1^{(\eta)}\overline{\X}^{i\ad}=0~,~~~%\hspace{38mm}
    \d_1^{(\L)}\overline{\X}^{i\ad}=-\frac{\e^{ij}}{M^2}~\D_{\r}\Big(\Dd_{\sd}\bar{\L}^{j\sd\ad}~W^{\r}\Big)~,
    \sn\\[6pt]
    \d_1^{(\eta)}\V=\frac{i~\e^{ij}}{2cM^2}
\Big[\eta^{j}_{(\b}{}^{\ad}\D^{\a}\Big(
        \Dd^{\gd}\D^{\b}\overline{\I}^{i}_{\a)\gd\ad}-\Dd^2\I^{i\b}{}_{\a)\ad}\Big)
+\frac{1}{3}\eta^{j\g\ad}\D^2\Big(2\Dd_{\ad}\overline{\I}^{i}_{\g}+\Dd^{\gd}\overline{\I}^{i}_{\g\gd\ad}\Big)
\Big],
    \d_1^{(\L)}\V=0~~~~~~~\sn
\end{IEEEeqnarray*}
Observe that $\d_{1}^{(\eta)}\H^{i}_{\a\ad}$ has a symmetrization between $\b$ and $\a$ indices.
If we expand the symmetrization and write explicitly the two terms we see that one of them
\Big(~$\D_{\a}\big[\frac{i}{2M^2}\e^{ij}\eta^{j\b}{}_{\ad}~W_{\b}\big]$~\Big) is just a free gauge
transformation \eqref{dLH} for some specific value of the gauge parameter
$\bar{L}_{\ad}\sim\frac{i}{M^2}\e^{ij}~\eta^{j\b}{}_{\ad}~W_{\b}$. Such terms of course correspond to just a
redefinition of the gauge parameter and we
will not include them in the set of non-trivial corrections to the gauge transformations.

Moreover, notice that $\d_{1}^{(\L)}\overline{\X}^{i}_{\ad}$ can be expanded by distributing the covariant
derivative giving two terms. One of these terms is
$\frac{1}{M^2}\e^{ij}\Dd^{\sd}\bar{\L}^{j}_{\sd\ad}~\E^{(\V)}$ where $\E^{(\V)}=\D^{\r}W_{\r}$. Of course,
this is nothing more than a trivial symmetry that our theory has. It is easy to check that the following
transformations preserve $S_0$:
\begin{equation}
    \d\overline{\X}^{i}_{\ad}=\frac{d}{M^2}~\e^{ij}~\Dd^{\sd}\bar{\L}^{j}_{\sd\ad}~\E^{(\V)}~,~
    \d\V=\frac{d}{cM^2}~\e^{ij}~\Dd^{\sd}\bar{\L}^{j}_{\sd}{}^{\ad}~\overline{\E}^{j~(\X)}_{\ad}~.
\end{equation}
Therefore, this part of the transformation does not correspond to non-trivial deformations of the free gauge
transformations and will also be ignored.

To conclude, the cubic interaction between a doublet of linearized non-minimal supergravity multiplets
($\Ysf=3/2$) and the vector multiplet ($\Ysf=1/2$) has the form
\begin{IEEEeqnarray*}{l}
    S_1=\frac{1}{M^2}\int d^8z~\e^{ij}\Bigg\{~\I^{j}_{\r(\b}{}^{\ad}~W^{\r}~\D^{\a}\Big(
        \Dd^{\gd}\D^{\b}\overline{\I}^{i}_{\a)\gd\ad} -\Dd^2\I^{i\b}{}_{\a)\ad}\Big)\n\\
        \hspace{39mm}+\frac{1}{3}~\I^{j~}_{\r}{}^{\g\ad}~W^{\r}~\D^2\Big(
    2~\Dd_{\ad}\overline{\I}^{i}_{\g}+\Dd^{\gd}\overline{\I}^{i}_{\g\gd\ad}\Big)~~\Bigg\}+c.c.
\end{IEEEeqnarray*}
This interaction is by construction manifestly invariant under transformations \eqref{dLV} and \eqref{dLH},
however invariance under the $\eta$ and $\L$-transformations require the following deformations:
\begin{IEEEeqnarray*}{l}\n
    \d_{1}\H^{i}_{\a\ad}=-\frac{i}{M^2}~\e^{ij}~\D^{\b}\Big(\eta^{j}_{\b\ad}~W_{\a}\Big)
    -\frac{1}{2M^2}~\e^{ij}~\Dd^{\sd}\bar{\L}^{j}_{\sd\ad}~W_{\a}~,\sn\\[6pt]
    \d_1\overline{\X}^{i}_{\ad}=-\frac{1}{M^2}~\e^{ij}~\D^{\r}\Dd^{\sd}\bar{\L}^{j}_{\sd\ad}~W_{\r}~,\sn\\[6pt]
    \d_1\V=\frac{i~\e^{ij}}{2cM^2}
\Big[\eta^{j}_{(\b}{}^{\ad}\D^{\a}\Big(
        \Dd^{\gd}\D^{\b}\overline{\I}^{i}_{\a)\gd\ad}-\Dd^2\I^{i~\b}{}_{\a)\ad}\Big)
    +\frac{1}{3}\eta^{j\g\ad}\D^2\Big(2\Dd_{\ad}\overline{\I}^{i}_{\g}
+\Dd^{\gd}\overline{\I}^{i}_{\g\gd\ad}\Big)\Big]\sn~.
\end{IEEEeqnarray*}
%%%%%%%%%%%%%%%%%%%%%%%%%%%%%%%%%%%%%%%%%%
%%%%%%%%%%%%%%%%%%%%%%%%%%%%%%%%%%%%%%%%%%
\section{Conclusions}
\label{sec:Outro}
Motivated by the success of the frame-like formulation in constructing consistent and non-trivial
interactions between various higher spin gauge fields, we develop a superspace first order formulation
for the linearized
supergravity and vector supermultiplets. This is done by relaxing the reality property of the corresponding
superfields in order to simplify the gauge transformation laws ($L$-transformation) and help us define simple invariants that are
used to write manifestly invariant actions. Of course, by doing that we introduce new, unwanted, degrees of
freedom which we eliminate by introducing a new local symmetry ($\eta$-transformation). Making the action invariant under this local
symmetry leads to the introduction of connection-like auxiliary superfields. In this case due to their
fermionic nature we have a pair of such `connections' $\W_{\b\a\ad}$ and $\vO_{\b\a\ad}$. After integrating
them out, we find a Lagrangian description for the $\Ysf=3/2$ multiplet in terms of a single
superfield $\I_{\b\a\ad}$ which packages together the complexified prepotential and its compensator
$\I_{\b\a\ad}=\D_{\b}\H_{\a\ad}+C_{\b\a}\overline{\X}_{\ad}$ (and $\I_{\ad}=C^{\b\a}\I_{\b\a\ad}$)
\begin{IEEEeqnarray*}{l}
    S_{(\Ysf=3/2)}[\I_{\b\a\ad}]=c\int
    d^8z~\Bigg\{\I^{\b\a\ad}~\Dd^2~\I_{\b\a\ad}+\frac{2}{9}~\I^{\b\a\ad}~\D_{\b}\Dd_{\ad}~\bar{\I}_{\a}~~+c.c.\\
        \hspace{47mm}        -\frac{8}{9}~\I^{\b\a\ad}~\D_{\b}\Dd^{\bd}~\bar{\I}_{\a\bd\ad}
        -2~\I^{\b\a\ad}~\Dd^{\bd}\D_{\b}~\bar{\I}_{\a\bd\ad}
    +\frac{4}{9}~\I^{\ad}~\D^{\a}\Dd_{\ad}~\bar{\I}_{\a}~~\Bigg\}~.
\end{IEEEeqnarray*}
By construction this action is  manifestly invariant under the $L$-transformation and also invariant under the
$\eta$-transformation. Moreover, $\I_{\b\a\ad}$ satisfies two Bianchi identities which give rise to
an additional emerging symmetry
($\L$-transformations) which matches what is expected from the prepotential description viewpoint.

We emphasize that the frame formulation in superspace has a long history. The full, non-linear theory of
supergravity was developed in superspace using supervielbein
and superconnection superfields. These superfields carry many additional degrees of freedom and one is forced
to impose several (conventional) constraints on them. After solving these constraints in terms of
prepotentials and linearizing the theory, all features of the first order formulation have been dissolved.
Our approach brings back some of these notions and not only allows us to construct cubic interactions
but also creates a path to bring the power of first order formulation to various higher spin supermultiplets.
This task will be completed in a following paper.

In this paper we apply this alternative description to construct cubic electromagnetic interaction of
supergravity. As always, fields that interact electromagnetically come in doublets and for this reason various
trivial symmetries emerge. These are symmetries that every field theory which involves many fields has, but
because they do not reduce the number of degrees of freedom they are not discussed often. In this case, trivial
symmetries can be used to show that the $3/2-3/2-1/2$ cubic vertex between two non-minimal superspin $\Ysf=3/2$
multiplets (linearized non-minimal supergravity) and one superspin $\Ysf=1/2$ multiplet (vector multiplet) must depend on the gauge invariant
superfield strength of the vector multiplet $W_{\a}$. This feature generalizes previously known results
for the non-supersymmetric
cubic vertex interactions of two spin 2 particles with a spin 1 particle.

Using Noether's procedure we find the first order correction to the action which describes the above
cubic interaction to be
\begin{IEEEeqnarray*}{l}
    S_1=\frac{1}{M^2}\int d^8z~\e^{ij}\Bigg\{~\I^{j}_{\r(\b}{}^{\ad}~W^{\r}~\D^{\a}\Big(
        \Dd^{\gd}\D^{\b}\overline{\I}^{i}_{\a)\gd\ad} -\Dd^2\I^{i\b}{}_{\a)\ad}\Big)\\
        \hspace{39mm}+\frac{1}{3}~\I^{j~}_{\r}{}^{\g\ad}~W^{\r}~\D^2\Big(
    2~\Dd_{\ad}\overline{\I}^{i}_{\g}+\Dd^{\gd}\overline{\I}^{i}_{\g\gd\ad}\Big)~~\Bigg\}+c.c.
\end{IEEEeqnarray*}
This cubic interaction requires non-trivial deformations of the $\eta$ and $\L$ gauge transformations
as follows:
\begin{IEEEeqnarray*}{l}
    \d_{1}\H^{i}_{\a\ad}=-\frac{i}{M^2}~\e^{ij}~\D^{\b}\Big(\eta^{j}_{\b\ad}~W_{\a}\Big)
    -\frac{1}{2M^2}~\e^{ij}~\Dd^{\sd}\bar{\L}^{j}_{\sd\ad}~W_{\a}~,\\[6pt]
    \d_1\overline{\X}^{i}_{\ad}=-\frac{1}{M^2}~\e^{ij}~\D^{\r}\Dd^{\sd}\bar{\L}^{j}_{\sd\ad}~W_{\r}~,\\[6pt]
    \d_1\V=\frac{i~\e^{ij}}{2cM^2}
\Big[\eta^{j}_{(\b}{}^{\ad}\D^{\a}\Big(
        \Dd^{\gd}\D^{\b}\overline{\I}^{i}_{\a)\gd\ad}-\Dd^2\I^{i~\b}{}_{\a)\ad}\Big)
    +\frac{1}{3}\eta^{j\g\ad}\D^2\Big(2\Dd_{\ad}\overline{\I}^{i}_{\g}
+\Dd^{\gd}\overline{\I}^{i}_{\g\gd\ad}\Big)\Big]~.
\end{IEEEeqnarray*}
These deformations can be mapped to the conventional superfield description according to the rule
\begin{equation}
    H^{i}_{\a\ad}=\H^{i}_{\a\ad}+\bar{\H}^{i}_{\a\ad}~,~\chi^{i}_{\a}=\X^{i}_{\a}~,~V=\V+\bar{\V}~,~
    \eta^{i}_{\a\ad}=\frac{i}{2}\big(\D_{\a}\bar{L}^{i}_{\ad}+\Dd_{\ad}L^{i}_{\a}\big)~.
\end{equation}

This interaction is of the type $\mathcal{L}_{1}\sim\Phi_1\Phi_2W_3$ as discussed in the introduction and can be
used as the starting point for higher spin generalizations of it. These interactions are known to exist and have
been constructed for non-supersymmetric theories and theories with on-shell supersymmetry. This work hopefully
provides the right set of tools needed in order to construct these interactions in superspace with manifest
supersymmetry.
Furthermore, it would be interesting to investigate whether the first order
description of supergravity given in \ref{subsec:superfofSG} can be pushed beyond the linearized limit and
make contact with the full non-linear theory. Also its applications to minimal formulations of
supergravity and higher spin multiplets should be studied. It is known that one can exchange
formulations of supergravity ---go from non-minimal supergravity ($n=-1$) to old-minimal supergravity
($n=-1/3$) and
vice versa--- by exploiting the duality between chiral and complex linear superfields. An example of employing
this duality can be found in \cite{Koutrolikos:2017qkx} where consistent interactions between matter and
higher spin supermultiplets are constructed. Using similar arguments one may find how to extend the above
results to a minimal formulation of supergravity.
%%%%%%%%%%%%%%%%%%%%%%%%%%%%%%%%%%%%%%%%%
%%%%%%%%%%%%%%%%%%%%%%%%%%%%%%%%%%%%%%%%%
\section*{Acknowledgments}
The work of I.~L.~B was partially supported by Ministry of Education of Russian Federation, project
FEWF-2020-0003. The work of K.~K. and S.~J.~G. is supported in part by the endowment of the Ford Foundation
Professorship of Physics at Brown University. Also K.~K. gratefully acknowledges the support of the Brown
Theoretical Physics Center.
\begin{multicols}{2}
{\small
\bibliographystyle{hephys}
\bibliography{references}
}
\end{multicols}
\end{document}